\documentclass[twocolumn,showpacs,pra]{revtex4}

\usepackage{amsmath, amsthm, amssymb}
\usepackage{graphicx}
\usepackage{dcolumn}
\usepackage{bm}

\newcommand{\de}[1]{\left( #1 \right)}

\newcommand{\refeq}[1]{Eq.~(\ref{#1})}

\renewcommand{\Re}[1]{{\mathrm{Re}}\de{#1}}

\newcommand{\ket}[1]{\left| #1 \right\rangle}
\newcommand{\bra}[1]{\left\langle #1 \right|}

\newcommand{\tr}{\mathrm{Tr}}

\newcommand{\identity}{\openone}

\newcommand{\eg}{{\it{e.g.~}}}
\newcommand{\ie}{{\it{i.e.~}}}
\newcommand{\etal}{{\it{et al.}}}

\newcommand{\be}{\begin{equation}}%{$$}
\newcommand{\ee}{\end{equation}}%{$$}
\newcommand{\bea}{\begin{eqnarray}}
\newcommand{\eea}{\end{eqnarray}}

\begin{document}

%\title{Bound entangled thermal states and the area law}
%\title{Distillability versus area laws in spin and harmonic-oscillator systems}
\title{Distillable entanglement and area laws in spin and harmonic-oscillator systems}

\author{Daniel Cavalcanti$^1$, Alessandro Ferraro$^1$, Artur Garc\'ia-Saez$^1$
and Antonio Ac\'\i n$^{1,2}$}%}\email{Daniel.Cavalcanti@icfo.es}
\affiliation{$^1$ICFO-Institut de Ciencies Fotoniques,
Mediterranean
Technology Park, 08860 Castelldefels (Barcelona), Spain\\%}
$^2$ICREA-Instituci\'o Catalana de Recerca i Estudis Avan\c cats,
Lluis Companys 23, 08010 Barcelona, Spain}

\begin{abstract}
  We address the presence of non-distillable (bound) entanglement in
  natural many-body systems. In particular, we consider standard
  harmonic and spin-$1/2$ chains, at thermal equilibrium and
  characterized by few interaction parameters.  The existence of bound
  entanglement is addressed by calculating explicitly the negativity
  of entanglement for different partitions.  This allows to
  individuate a range of temperatures for which no entanglement can be
  distilled by means of local operations, despite the system being
  globally entangled. We discuss how the appearance of bound
  entanglement can be linked to entanglement-area laws, typical of
  these systems. Various types of interactions and topologies are
  explored, showing that the presence of bound entanglement is an
  intrinsic feature of these systems. In the harmonic case, we
  analytically prove that thermal bound entanglement persists for
  systems composed by an arbitrary number of particles.  Our results
  strongly suggest the existence of bound entangled states in the
  macroscopic limit also for spin-$1/2$ systems.
\end{abstract}


\maketitle

%%%%%%%%%%%%%%%%%%%%%%%%%%%%%%%%%%%%%%%%%%%%%%%%%%%%%%%%%%%%%%%
\section{Introduction}

An ubiquitous scenario in Quantum Information Science consists of two
(or more) parties sharing entangled quantum states and performing
operations on them. Ideally the shared states are pure and maximally
entangled. However, this is hardly the case in practice. As a matter
of fact, the available states are always mixed, due to unavoidable
errors in the preparation stage or noise in the considered process.
Entanglement is also very fragile in this sense, and this is certainly
the main obstacle to actual applications of Quantum Information ideas.

Entanglement distillation plays an important role to correct the
degradation of entanglement in real situations \cite{Benn1,Benn2}. It
consists on the application of local operations supplied by classical
communication (LOCC) that allow the parties to extract maximally
entangled states out of a bunch of mixed states. Unfortunately not all
entangled states are distillable \cite{Hor}. There are states $\rho$
from which no LOCC strategy is able to extract pure-state entanglement,
even if many copies of $\rho$ are available. These states are known as
\emph{bound entangled}.
%
%One of the greatest turning points in quantum
%theory was the understanding that quantum correlations can be
%treated as valuable resources. In this context many quantum
%information tasks are based on several distant observers sharing
%pure entangled states and performing quantum operations on them.
%However, in practical situations, environmental noise (often in
%form of thermal noise) is always present and then we unavoidably
%deal with mixed states. In the bipartite scenario, the first
%attempt to overcome this degradation of entanglement led to the
%idea of entanglement distillation \cite{Benn1,Benn2}. A
%distillation protocol is a sequence of local operations assisted
%by classical communication (LOCC) which allows one to transform
%$N$ copies of a mixed entangled state into $M<N$ copies of a pure
%maximally entangled state. These distilled states can be
%subsequently used for quantum information purposes. A question
%rapidly comes up: can one always find a distillation protocol for
%an arbitrary entangled mixed state? Curiously the answer is no:
%there are entangled states for which no LOCC protocol succeeds in
%extracting pure entanglement from them. These states are known as
%bound entangled states \cite{Hor}.

%After the discovery of bound entangled states it became natural to
%look for examples of such a states.
The existence of bound entangled states was proven in~\cite{Hor} by noting that any quantum state with a positive partial transposition (PPT) \cite{Peres} is non-distillable. Since there were already examples of PPT entangled states~\cite{PH}, the connection between non-distillability
and the positivity of partial transposition automatically led to the first examples of bound entangled states.
%Thus if one finds a non-separable state with
%positive partial transposition (PPT) it must be bound entangled .
Later, strong evidence was provided for the existence of
bound entangled states with non-positive partial transposition (NPPT) \cite{NPPT}. Independently of partial transposition, bound entangled states have a clear operational definition: an entangled state of $n$ parties is bound entangled
whenever the $n$ parties cannot distill any pure-state
entanglement out of it by LOCC. Although several examples of bound
entangled states have appeared so far \cite{HorReview}, simple
recipes to construct such kind of states are still lacking.
%Then
%up to now, a question remained open: do bound entangled states
%appear inherently in nature? In particular, do these states emerge
%in standard quantum many-body systems?
It is thus not completely clear whether bound entanglement is basically a
mathematical artifact or it ``naturally" appears in physically relevant quantum systems. The main motivation of this work goes precisely along this direction, since we study whether bound entanglement is present in standard many-body quantum systems. By this, we mean systems {\it i)} characterized by a few interaction parameters, {\it ii)} in thermal equilibrium with their environment, {\it iii)} consisting of a macroscopic number of particles.

Very recently a few works have also addressed this question. In
Ref.~\cite{Geza} bound entanglement was detected in the thermal state
of spin systems, consisting of up to $9$ spins. More recently,
three-qubit bound entangled states were obtained as the reduced state
of the $XY$ model in the thermodynamical limit \cite{patane}. In a
previous contribution we have considered the problem of finding bound
entangled thermal states in some specific system consisting of a
macroscopic number of harmonic oscillators \cite{nos}. The main goal
of the present article is to give a detailed description of these
results and extend them to other systems.
%
%In this paper we show that bound entangled states appear naturally as
%thermal states of several many-body systems, described both by
%continuous and discrete variables.
Specifically, we directly calculate the entanglement in different
partitions of thermal states for harmonic oscillators and spin systems
and identify a temperature range for which bound entanglement is
present. In the harmonic case with nearest neighbor interaction, we first
consider systems composed of hundreds of particles and then explicitly
prove that bound entanglement indeed persists in the macroscopic
limit, \ie when $n\rightarrow\infty$. Moreover we offer an explanation
of the results in terms of
%The fundamental intuition to elucidate these
%results comes from
the entanglement area law, a property satisfied by standard many-body systems \cite{EntMBS}.
%satisfies and  by  according to
%which the entanglement between two regions scales as the surface
%separating them .
%In this sense, the route
%we take here goes in the reverse direction with respect to the one
%usually pursued by many recent works \cite{QPT}: we borrow concepts
%from the condensed matter field to get new insight on quantum
%information science.
Due to the general validity of area laws, we see that the presence
of bound entanglement is a common feature of these
systems. We explicitly analyze a variety of different systems, ranging
from critical to non-critical situations and to spin chains
characterized by a complex behavior of the entanglement in the ground
state. In all these cases we see that there is a range of temperatures
for which no entanglement can be distilled by means of LOCC, despite the system being globally entangled.

The paper is organized as follows. In the next Section we
introduce the main intuition that elucidates the emergence of bound
entanglement in thermal states, recalling briefly some known results
about entanglement-area laws. Then, in Sec.~\ref{ho} we present
the analysis of harmonic systems. In particular, in
Sec.~\ref{numerics} numerical results regarding finite size systems
are presented, whereas in Sec.~\ref{analytics} we give the
analytical calculations in the macroscopic limit for systems with
nearest neighbor interactions. In Sec.~\ref{spin}, we show that
similar results can also be obtained for spin systems, even if in this
case, and due to numerical limitations, systems composed only by a small number of particles have been considered. We close the paper with some concluding remarks in Sec.~\ref{esco}.

%%%%%%%%%%%%%%%%%%%%%%%%%%%%%%%%%%%%%%%%%%%%%%%%%%%%%%%%%
\section{Multipartite Bound entanglement and area laws}\label{area}
Let us present the main intuition behind our construction of bound
entangled thermal states. Consider a quantum system of $n$
particles
%(labeled from $1$ to $n$)
described by a local Hamiltonian. For the sake of simplicity, in what
follows we restrict our analysis to one-dimensional systems of $n$
particles. A common property of these systems is that the ground-state
entanglement obeys an area law. This means that the entanglement of a
bipartite splitting of the system into two groups scales at most as
the number of connections between the groups (i.e. the area which
separates them). We recall that the entanglement for a bipartite
splitting of pure states is uniquely quantified by the entropy of one
of the subsystems \cite{Ben96,PR97}. In general, this quantity scales
as the volume of the subsystem, and not as its boundary, for an
arbitrary state belonging to the whole Hilbert space of the $n$
particles. However, as said, this is not the case for the great
majority of ground states of local Hamiltonian systems.

%
%Notice that typically states in Hilbert space exhibit a volume
%scaling, so such a behavior is a priori surprising and it can be
%connected to the emergence of a characteristic correlation length
%in the system \cite{areath}.
%
Entanglement-area relations were first recognized in connection
with the physics of black holes, for which the Von Neumann entropy
scales as the surface at the event horizon~\cite{bh}. In recent
years, the advances of entanglement theory allowed for the
exploration of this behavior also for a variety of systems typical
of condensed matter physics (see Ref.~\cite{EntMBS} and references
therein). Remarkably, many analytical results have been found for
harmonic systems for which concepts from the theory of Gaussian quantum
states can be applied. As an example, a strict entanglement-area
relation has been established for non-critical one-dimensional
systems with finite range interaction, while logarithmic
corrections appear in the critical case \cite{Pl,Unanyan,Area}. In
particular, the entanglement, as measured by the log-negativity
\cite{VidWer}, has been proved to scale proportionally to the
area. Similar behaviors have also been found for spin systems.
Concerning thermal states, it has recently been demonstrated that
the total amount of correlations (measured by the mutual
information) in a bipartite split scale at most as the area
\cite{areath}. This, in turn, gives an upper bound to the
entanglement.

Consider now a translationally invariant system composed by an even
number $n$ of particles (labeled from $1$ to $n$). Let us focus our
attention on two different partitions of it, one in which a contiguous
half of the particles belongs to group $A$ and the other half to $B$
(we will refer to such kind of partition as half-half), and another
partition in which the particles with even label belong to $A$ and the
rest to $B$ (even-odd partition). Because of the area law, the
entanglement saturates for sufficiently large $n$ for the
half-half partition, while it increases with $n$ for the even-odd
partition. In this configuration, it is reasonable to expect that, by
increasing the temperature, the entanglement in the even-odd partition
is more robust to thermal noise than in the half-half partition, and
that this behavior is preserved for large systems.  Denote by
$T^{h:h}_{\rm th}$ ($T^{e:o}_{\rm th}$) the threshold temperatures at
which the partial transposition with respect to all half-half
(even-odd) partitions becomes positive \cite{note1}.  Because of the
area law, one can expect that $T^{h:h}_{\rm th}$ is strictly smaller
than $T^{e:o}_{\rm th}$.  Thus, it emerges a range of temperatures for
which the system is still entangled (as detected by the entanglement
in the even-odd partition), nevertheless single particles cannot
distill pure entanglement (as the half-half partitions become PPT).
This is because, for any pair of particles, there is always a
half-half partition for which they are in opposite sides and the
partial transposition is positive according to this splitting
(remember that this is a sufficient condition for non-distillability
\cite{Hor}). In other words, bound entangled states are expected to appear under these conditions. The rest of the paper is devoted to
verify and put on solid grounds this intuition for various relevant examples of many-body systems.

Before proceeding, let us relate our finding to two related works
studying the presence of bound entanglement in systems with local
interactions. First, in Ref.~\cite{Geza} the existence of thermal
entangled states which are PPT with respect to any bipartition was
proven for systems consisting of up to 9 spins. For such states,
distillation of pure-state entanglement is impossible even if the
parties arbitrarily combine together. As said above, we will instead
show the existence of bound entangled states only with respect to
fully local distillation procedures but for an arbitrary number of
particles. Actually, for the harmonic systems considered here we
explicitly prove that no entangled states PPT with respect to any
bipartition exist. Second, in Ref.  \cite{patane}, the authors find
bound entanglement in the reduced three-qubit state of a macroscopic
system. Here, we focus on the existence of bound entanglement in the
whole thermal state of the system.

%%%%%%%%%%%%%%%%%%%%%%%%%%%%%%%%%%%%%%%%%%%%%%%%%%%
\section{Harmonic oscillators}\label{ho}
In this Section, we first introduce the harmonic systems that we
are going to consider in order to test the ideas exposed above.  Then,
in Sec. \ref{numerics}, we perform numerical calculations for
different types of interactions in systems of finite size. In Sec.
\ref{analytics}, analytical results are presented for the
significant case of nearest-neighbor interaction in the macroscopic
limit.
\subsection{The systems}
Consider a system composed of $n$ harmonic oscillators, each one
associated with position and momentum operators $x_i$ and $p_i$
respectively ($i=1,\dots,n$), described by the Hamiltonian
\begin{equation}\label{Hosc}
H=\frac{1}{2}\left(\sum_i p_i^2+\sum_{i,j}x_i V_{i,j} x_j\right).
\end{equation}
The diagonal elements of the matrix $V$ describe the on-site
interaction, while the non-diagonal terms give the coupling between
oscillators $i$ and $j$. Note that this Hamiltonian is quadratic in
the canonical coordinates and the oscillators are coupled through
their position degrees of freedom. In this scenario both the ground
and the thermal states turn out to be Gaussian, so they are
completely described by their covariance matrix $\gamma$. Introducing the
vector $S=(x_1,\dots,x_n,p_1,\dots,p_n)$, the latter is defined as follows
\begin{equation}
\gamma_{kl}=\Re{\tr\{\varrho[S_k-\bar S_k][S_l-\bar S_l]\}}
\end{equation}
where $\varrho$ denotes the density matrix of the state and $\bar
S_k=\tr(\varrho S_k)$. Considering the thermal state
$\varrho=\exp[-H/T]/\tr\{\exp[-H/T]\}$ at temperature $T$, the
corresponding covariance matrix is given by \cite{Aud}:
\begin{equation}
\gamma(T)=[V^{-1/2}W(T)]\oplus[V^{1/2}W(T)]\,,
\end{equation}
where
\begin{equation}
W(T)=\identity_n +2[\exp(V^{1/2}/T) -\identity_n]^{-1}
\end{equation}
and $\identity_n$ denotes the $n\times n$ identity matrix. In the
ground-state case $W(0)$ is given by the identity matrix and thus
$\gamma(0)=V^{-1/2}\oplus V^{1/2}$.

The entanglement properties of the ground and
thermal states corresponding to Hamiltonian \eqref{Hosc} were first
studied by Audenaert \etal\ \cite{Aud}. There, an analytical expression for
the entanglement (quantified by the log-negativity \cite{VidWer})
between two complementary groups of oscillators, $A$ and $B$, was
given in terms of the covariance matrix of the state, which can be
written, in turn, only in terms of the matrix $V$. Then one gets the
general formula for the log-negativity of a thermal state at temperature $T$:
\begin{equation}
E_l=\sum_{k=0}^{n-1} \log_2
\{\max[1,\lambda_k(Q)]\},
\label{logneg}
\end{equation}
where $Q=P\,\omega^-\, P\omega^+$ and
$\omega^\pm=W(T)^{-1}V^{\pm\frac{1}{2}}$. We denoted by
$\{\lambda_k[Q]\}_{k=0}^{n-1}$ the spectrum of the matrix $Q$, whereas
$P$ is an $n\times n$ diagonal matrix with the $i$-th entry given by
$1$ or $-1$ depending on which group, $A$ or $B$, oscillator $i$
belongs to. This study was later extended in Ref.~\cite{Pl} where an
area law for ground-state entanglement was proven. As far as for
thermal states, an upper bound for the entanglement in terms of the
number of connecting points in a given bipartition was also
established \cite{Area}, but the lack of a lower bound limits us to
get a strict area law in this case.

In the case of a harmonic chain with nearest-neighbor
interactions and periodic boundary conditions, the system is
described by the Hamiltonian \eqref{Hosc} with a circulant potential
matrix $V$ given by
\begin{equation}\label{Vnn}
V^{\rm n}={\rm circ}(1,-c,0,\dots,0,-c)\,.
\end{equation}
The system is defined for $0\le c<1/2$ and it is equivalent to a
chain of harmonic oscillators coupled with a spring-like interaction.
A variety of physical systems can be modeled by such interaction,
going from vibrational degrees of freedom in crystal lattices and ion
traps to free scalar Klein-Gordon field. We recall that its
ground-state entanglement exhibits a critical behavior when
$c\rightarrow 1/2$ \cite{Botero,Pl,Schuch}.

In order to test the general validity of the connection between bound
entanglement and area laws mentioned in Sec.~\ref{area}, we have
considered also other systems. As an example we will report here some
results concerning the \emph{next-to-nearest interaction} defined by
the potential matrix \cite{Unanyan}
\begin{equation}\label{Vuf}
V^{\rm nn}={\rm circ}(2+4\mu^2,-4\mu,1,0,\dots,0,1,-4\mu)\,.
\end{equation}
This system has been shown to be critical (gapless) and to violate the
area law for $0<\mu<1$ in the ground state. Nevertheless, as already
mentioned, no violation of the area law is possible for non-zero
temperatures in view of the results in Ref.~\cite{areath}.
%%%%%%%%%%%%%%%%%%%%%%%%%%%%%%%%%%%%%%%%%%%%%%%%%%%%%%%%%%%
\subsection{Numerical results}\label{numerics}
\begin{figure} {\includegraphics[width=0.23\textwidth]{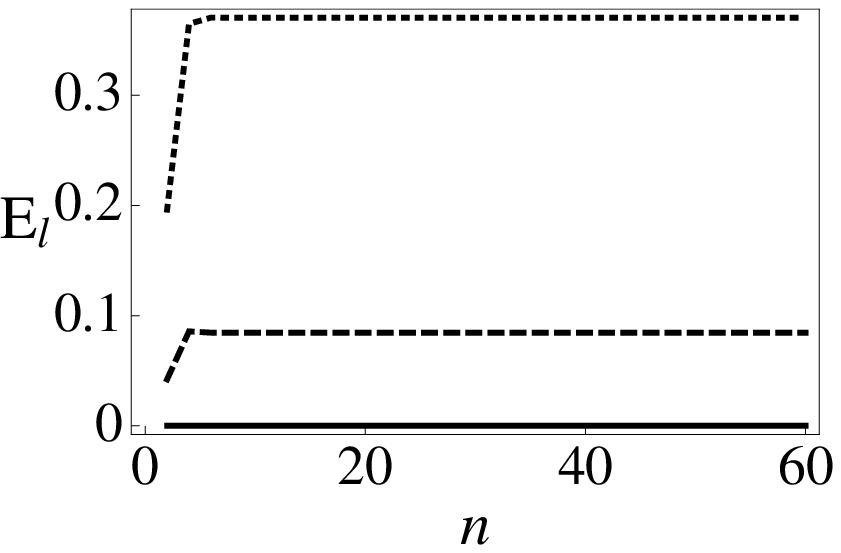}}
 {\includegraphics[width=0.22\textwidth]{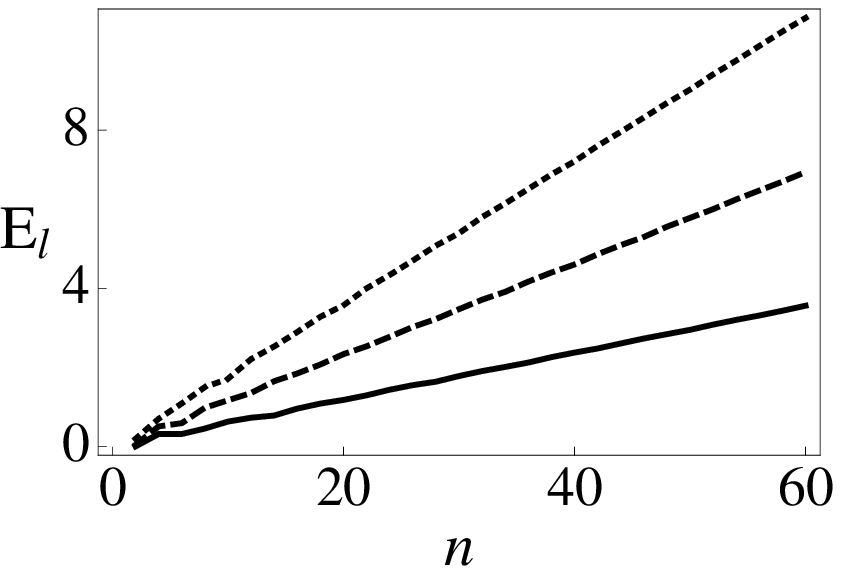}}

 \caption{Log-negativity $E_l$ as a function of the number of
   oscillators $n$ for half/half (left panel) and the even/odd (right
   panel) partitions. Temperatures are $T=0.45$ (solid line), $T=0.4$
   (dashed line), and $T=0.35$ (dotted line), and $c=0.4$ (the same
   behavior is found for other values of $c$). For $T=0.45$ the state
   is bound entangled.}
\label{area_nn}\end{figure}

Let us first consider the nearest-neighbor interaction given by
\refeq{Vnn}. We used \refeq{logneg} to compute the log-negativity for
the even-odd and the half-half partitions \cite{note1} for different
temperatures and number of particles.  As already recalled, in
Ref.~\cite{Aud} it is shown that a strict area law holds for this
system in the ground state. Actually, our calculations show that the
log-negativity follows a strict area law for non-zero
temperatures as well. We depicted in Fig.~\ref{area_nn} the log-negativity as a function of the number of particles, for fixed coupling and
different temperatures. One can clearly see that, apart from a
transient for small $n$, the log-negativity increases linearly with
$n$ for the even-odd case, while it saturates for the half-half
partition. Furthermore, the change of the system temperature just
affects the rate at which entanglement increases with $n$ for the even-odd
partition and the entanglement saturation value for the half-half
partition. The validity of a strict area law has then a remarkable
consequence: being the log-negativity dependence on $n$ the same (linear or constant) for all $T$, then $E_l$ goes to zero as $T$ increases independently of $n$. In other words, the threshold temperatures $T^{h:h}_{\rm th}$ and $T^{e:o}_{\rm th}$ should be independent of the size of the system (apart for small $n$, for which the area law shows a transient). Regarding the presence of bound entanglement, we see in
Fig.~\ref{area_nn} that, \eg, for $T=0.45$ the log-negativity in the
half-half partition is zero, meaning that single particles cannot
distill, nevertheless the system is entangled, as shown by the
non-zero log-negativity in the even-odd partition. For the reasons
exposed in Sec.~\ref{area} we then conclude that for that temperature
the system is bound entangled.

\begin{figure} {\includegraphics[width=0.5\textwidth,height=0.3\textwidth]{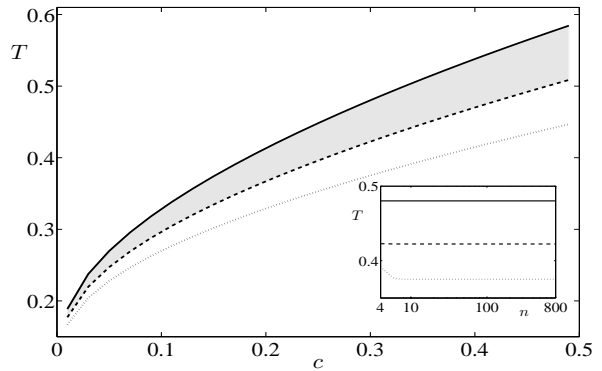}}
  \caption{Threshold temperatures above which the log-negativity is
    zero in the even-odd ($T^{e:o}_{\rm th}$, solid line) and
    half-half ($T^{h:h}_{\rm th}$, dotted line) partitions. The
    thresholds are plotted as a function of the coupling constant $c$
    for the Harmonic chain with nearest-neighbor interactions composed
    by 800 oscillators.  The intermediate line (dashed) shows the
    temperature $T_{\rm th}^{1:n-1}$ above which the state is
    separable in all the partitions $1:n-1$.  {\bf{Inset:}}
    $T^{e:o}_{\rm th}$ (solid line), $T^{h:h}_{\rm th}$ (dotted line),
    and $T_{\rm th}^{1:n-1}$ (dashed line) as a function of the number
    $n$ of oscillators composing the system (log-lin scale).  The
    oscillators interact via nearest-neighbors couplings with $c=0.3$
    (the same behavior is found for other values of $c$).  The gap
    $T^{e:o}_{\rm th}-T^{h:h}_{\rm th}$ is seen to remain constant
    with the size of the system, apart from an initial transient.}
\label{PhD}\end{figure}

An exhaustive evidence of the presence of bound entanglement for
any coupling constant is given in the $T-c$ diagram of
Fig.~\ref{PhD}. There we see that the threshold $T^{e:o}_{\rm th}$ is
strictly larger than $T^{h:h}_{\rm th}$ for any $c$, indicating the
existence of a bound entanglement region as soon as $c\neq0$. The
calculations are in this case performed for systems composed by
800 oscillators. The range of temperatures
$T^{h:h}_{\rm th}<T<T^{e:o}_{\rm th}$ for which bound entanglement is
guaranteed is seen to increase when the system approaches the
critical point $c=0.5$.

As said, the fact that all the half-half partitions are PPT
implies that no entanglement can be distilled by means of local
operations performed on each particle. However, it does not exclude the
possibility that each oscillator is entangled with the others
\cite{Smolin}. Interestingly, we also found a region of
temperatures for which none of the oscillators is entangled with
the rest of the chain, yet the global state is entangled (shaded
region in Fig.~\ref{PhD}). This is shown by calculating the
log-negativity for the partitions $1:n-1$, \ie one particle versus
the others. Since the partition consists of $1$ versus $n-1$
modes, the PPT condition turns out to be sufficient for
separability \cite{1mode}. We have calculated also the threshold
temperatures corresponding to a generic contiguous splitting
$\frac{n}{2}-m:\frac{n}{2}+m$ (with $0\le m \le \frac{n}{2}-1$).
It can be seen that these threshold temperatures increase with $m$, that is, the entanglement in the bipartition $1:n-1$ is
the more robust to thermal noise among all these splittings.

In order to analyze the dependence of the results above on the
system size, we have plotted the threshold temperatures as
a function of $n$ (see the inset of Fig.~\ref{PhD} ). In
accordance to what was discussed, we see that the threshold
temperatures remain constant.  As a consequence, also the gap
$T^{e:o}_{\rm th}-T^{h:h}_{\rm th}$, which determines the temperature
range in which bound entanglement is present, is independent of the
size of the system. These results strongly suggest that bound
entanglement can also be observed in the macroscopic limit, an
issue on which we will come back in the next session.

\begin{figure} {\includegraphics[width=0.5\textwidth,height=0.3\textwidth]{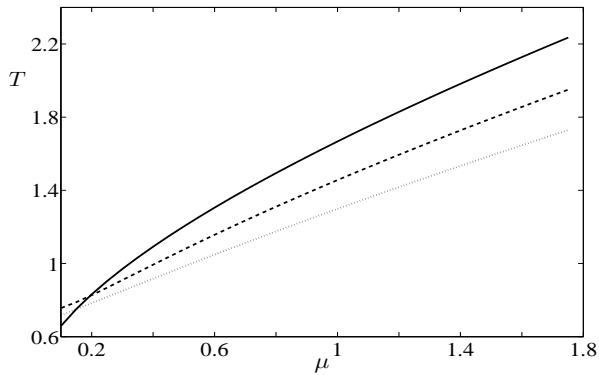}}
\caption{$T^{e:o}_{\rm th}$ (solid line), $T^{1:n-1}_{\rm th}$ (dashed
line),
  and $T^{h:h}_{\rm th}$ (dotted line) as a function of the coupling
  constant $\mu$ for the Harmonic chain with next-nearest-neighbor
  interactions in \refeq{Vuf} composed by 200 oscillators.}
\label{PhDuf}\end{figure}

Due to the generality of the area law, the results obtained for the
nearest-neighbor model above are expected to be valid in a variety of
different cases. As an example we report here the next-to-nearest
interaction $V^{\rm nn}$ introduced in the previous subsection. In the
$T-\mu$ diagram of Fig.~\ref{PhDuf} we can see that there is again a
wide range of temperatures for which bound entanglement is present, as
detected by the coexistence of nonzero log-negativity in the even/odd
partition and zero log-negativity in the half/half one.  Notice that
the absence of an energy gap for $\mu\le1$ does not affect
qualitatively the presence of bound entanglement \cite{note2}. As said
above, this can be related to the fact that an area law also holds for
gapless systems in case of non-zero temperatures.

\subsection{Analytical results}\label{analytics}

\begin{figure} {\includegraphics[width=0.5\textwidth,height=0.3\textwidth]{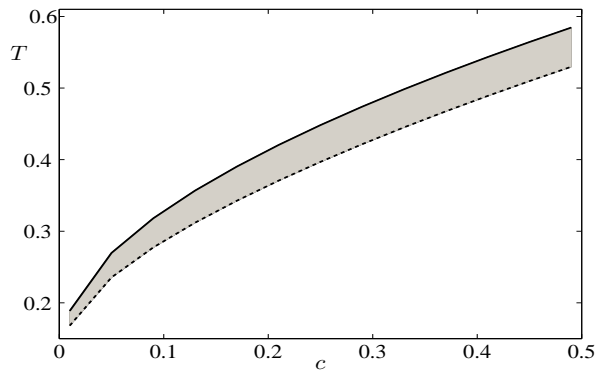}}
\caption{The solid line represents $T^{e:o}_{\rm th}$ in the
macroscopic
  limit, accordingly to Ineq.~(\ref{negeotlim}): below this line the
  state of the system is entangled. The dashed line represents the
  upper bound to $T^{h:h}_{\rm th}$ as given by Ineq.~(\ref{Thh}) (e.g.,
  we considered $m=10$ and $s=3$): above this line the log-negativity
  in the half-half partition is zero. In the shaded region then we can
  guarantee the presence of bound entanglement in the macroscopic
  limit.}
\label{tlim_phd}\end{figure}

As pointed out in the previous section, the validity of a strict area
law suggests both the independence of the threshold temperatures from
the size of the system as well as the preservation of thermal bound
entanglement for macroscopic systems. In this section we will
analytically prove that this is actually the case. In particular, we
will establish, for the paradigmatic system in \refeq{Vnn}, a formula
for $E_l$ in the even-odd partition and an upper bound for
$T^{h:h}_{\rm th}$ in the half-half partition. Both will be obtained
in the macroscopic limit, \ie for $n\rightarrow\infty$. The remaining
part of this Section is quite technical. The reader not interested in
the details can skip the proof, whose results are for convenience
summarized in Fig.~\ref{tlim_phd}.

Let us start with the case of the even-odd partition, for which we can
analytically prove the linear increase of the log-negativity with the
system size, as well as obtain the threshold temperature $T^{e:o}_{\rm
  th}$ as a function of the coupling $c$. The proof involves the
calculation of the spectrum of $Q$ in \refeq{logneg}, and
it follows closely the one given in Ref.~\cite{Aud} for the ground-state case. Consider the Hamiltonian \eqref{Hosc} and a generic
potential with translational invariance symmetry and periodic boundary
conditions. Then, $V$ is a circulant matrix and the matrices
$\omega^\pm$ can be diagonalized by a discrete Fourier transformation
implemented by \begin{equation} \Omega_{k,l}=\exp\left(kl \frac{2\pi
      i}{n}\right)/\sqrt{n},
\end{equation}
with $0\le k,l\le n-1$.  Namely, one has that
$\Omega\omega^\pm\Omega^\dagger=D^\pm$ with
$D^\pm_{k\,l}=\delta_{k\,l}\,d^\pm_k$ and
$d^\pm_k=\Lambda^{\pm1/2}_k\tanh(\sqrt{\Lambda_k}/2T)$ where
$\Lambda_k$ are the eigenvalues of $V$. Concerning the matrix $P$, one has
that:
\begin{equation}
\Omega\,P\Omega^\dagger=\tilde{P}=\left(\begin{array}{cc} 0 & \identity_{n/2} \\
    \identity_{n/2} & 0\end{array}\right)\;.
\end{equation}
The spectrum of $Q$ coincides then with the one of
$\tilde{P}D^-\tilde{P}D^+$, which in turn is straightforward to
calculate due to its block diagonal structure. The eigenvalues of $Q$
are hence given by $d^-_k d^+_{k+n/2}$ and $d^-_{k+n/2} d^+_k$, for
$k=0,\dots,n/2-1$.  Specializing these results to the nearest-neighbor
case in \refeq{Vnn}, one has that the eigenvalues of $V^{\rm n}$ are given by
$\Lambda_k=1-2c\cos{\left(2\pi k/n\right)}$. Defining the function
\begin{equation}
f(k,n,c,T)=\sqrt{\frac{\Lambda_{k+n/2}}{\Lambda_{k
      }}}
\tanh\left(\frac{\sqrt{\Lambda_k}}{2T}\right)
\tanh\left(\frac{\sqrt{\Lambda_{k+n/2}}}{2T}\right) \label{ff}
\end{equation} one has that the eigenvalues of $Q$ that can
contribute to the log-negativity (\ie\!, that can be larger than
one in dependence of the parameters) are given by $f(k,n,c,T)$,
with double multiplicity and $k=0,\dots,n/4$ (for $n$ multiple of
$4$, $n\ge4$). The function $f(k,n,c,T)$ is non increasing in $k$
for the range of interest ($0\le c<1/2$ and $T\ge0$), and in
particular it attains its maximum for $k=0$. As a consequence, the
log-negativity of the even-odd partition is different from zero
when the temperature $T$ is such that $f(0,n,c,T)>1$, namely:
\begin{equation} \sqrt{\frac{1+2c}{1-2c}}
\tanh\left(\frac{\sqrt{1-2c}}{2T}\right)\tanh\left(\frac{\sqrt{1+2c}}{2T}\right)>1
\label{negeotlim} \end{equation}
In particular, the curve
$f(0,n,c,T)=1$ gives the threshold temperature $T^{e:o}_{\rm th}$. It
coincides with the one depicted in Fig.~\ref{PhD} (solid line) and
it is worth noticing that it does not depend on the total number
of particles $n$ (see also the solid line in the inset of
Fig.~\ref{PhD}), \ie it holds also in the macroscopic limit. For
temperatures below $T^{e:o}_{\rm th}$ there exists a
$\overline{k}(n,c,T)$ such that $f(k,n,c,T)>1$ for
$k<\overline{k}(n,c,T)$, which in turn gives rise to the following
expression for the log-negativity:
\begin{equation}
E_l=\sum_{k=0}^{\overline{k}(n,c,T)} \log_2 \{f(k,n,c,T)\}.
\end{equation}
Following again Ref.~\cite{Aud}, we have that for large $n$ the
sum over $k$ can be replaced by an integral over $x=2\pi k/n$ :
\begin{equation}
E_l=\frac{n}{2\pi}\int_{0}^{\overline{x}(c,T)} dx\log_2
\{f(x,c,T)\}.
\end{equation}
Since $f(k,n,c,T)$ in Eq.~(\ref{ff}) actually depends only on the
parameter $x$, we have that both the integrand and the upper bound
$\overline{x}(c,T)$ turn out to be independent of $n$. As a
consequence the log-negativity grows linearly with the system size
in the macroscopic limit also for non-zero temperatures.

Regarding the half-half partition, we were not able to find an
analytic expression for the log-negativity in the macroscopic
limit. Nevertheless it is possible to find an exact upper bound
for the threshold temperature $T^{h:h}_{\rm th}$. As said, this
allows to identify an interval of temperatures for which the
presence of bound entanglement can be guaranteed also in the
macroscopic limit. We proceed as follows. Let us define the matrix
\begin{equation} X_{ij}=\omega^-_{ij}\sum_{k=0}^{n/2-1}\sum_{h=n/2}^{n-1}
(\delta_{ik}\delta_{jh}+\delta_{jk}\delta_{ih})\,
\end{equation}
where $\delta_{ij}$ denotes the Kronecker delta. According to
Ref.~\cite{Area}, the log-negativity in the half-half partition is
zero when the following inequality is satisfied:
\begin{equation}
\lambda_{\rm min}[W(T)]^{-2}+2\,{\rm
max}_i|\lambda_i[X\omega^+]|<1\;, \label{cineq}
\end{equation}
where the minimum eigenvalue of $W(T)$ is given by:
\begin{equation}
\lambda_{\rm
min}[W(T)]=\frac{e^{\sqrt{1+2c}/T}+1}{e^{\sqrt{1+2c}/T}-1}.
\label{lmin}
\end{equation} Recognizing that the second term in
the left hand side of Ineq.~(\ref{cineq}) is twice the spectral
radius $r(X\omega^+)$ of the matrix $X\omega^+$, we can use any
matrix norm to bound it from above \cite{HJ}. In particular, let
us consider the maximum row sum matrix norm $||\cdot||_\infty$,
defined for a generic matrix $A$ as:
\begin{equation}
||A||_\infty\equiv{\rm max}_i\sum_j |A_{ij}|\; \label{ninf}
\end{equation}
(in what follows, we will omit the subscript for the sake of
convenience). An upper bound for $r(X\omega^+)$ is then given by
$r(X\omega^+)\le||X\omega^+||$, from which it follows that:
\begin{equation}
 r(X\omega^+)\le||X||||\omega^+||\;. \label{rbound}
 \end{equation}
Before proceeding in bounding $||X||$ and $||\omega^+||$, let us
write explicitly their elements in the macroscopic limit. Recall
that $\omega^\pm$ are circulant matrices, hence completely
specified by their first row $\omega^\pm={\rm
circ}(v_0^\pm,\dots,v_{n-1}^\pm)$, whose elements are in turn
given by the discrete Fourier transform of the eigenvalues
$d_k^\pm$ introduced above:
\begin{equation}
v_l^\pm=\frac1n
\sum_{k=0}^{n-1}d_k^\pm \exp \left(\frac{2\pi i}{n}k l \right) \;.
\end{equation}
In the limit $n\rightarrow\infty$ the discrete sum above
yields the Riemann sum for the following integral in the argument
$x=2 \pi k/n $: \begin{equation}
v_l^\pm=\frac{1}{2\pi}\int_0^{2\pi}dx\,d^\pm(x) e^{i x l}\;,
\label{vls} \end{equation} that is, $v_l^\pm$ are the Fourier
coefficients of the periodic functions $d^\pm(x)$:
\begin{equation} d^\pm(x)=(1-2c\cos
x)^{\pm1/2}\tanh(\frac{\sqrt{1-2c\cos x}}{2T})\,.
\end{equation}
Notice that $d^\pm(x)$ are smooth functions, which implies an
exponential decay of $|v_l^\pm|$ as a function of $l$. In fact, for
any integer $s$, by integrating by parts $s$ times we have:
\bea\label{decay} |v_l^\pm|&=&\frac{1}{2\pi} \left|\frac{1}{(i
    l)^s}\int_0^{2\pi}dx\, e^{i x l}\frac{d}{dx^s}d^\pm(x)\right|
\nonumber \\
&\le&\frac{1}{2\pi} \frac{1}{l^s} C_s^\pm \eea where
$C_s^\pm\equiv\int_0^{2\pi}dx\, \left|\frac{d}{dx^s}d^\pm(x)\right|$.
Let us now bound first $||\omega^+||$. Being $\omega^+$ a circulant
matrix, there is no need to look for the maximum in (\ref{ninf}), as
the rows are composed by the same elements. Then one can write, for
any integer $m$, $||\omega^+||=S^+_m+{\cal E}_m^+$, where we defined
the partial sum and the residual term, respectively, as follows:
\begin{equation}
S^+_m\equiv\sum_{l=-m}^{m}\!\!|v_l^+|\;,\qquad
{\cal E}_m^+\equiv2\sum_{l=m+1}^{\infty}\!|v_l^+|\;.
\end{equation}
In order to obtain a bound on $||\omega^+||$ one can now fix $m$,
calculate explicitly $S^+_m$ and bound ${\cal E}_m^+$ from above.
This latter step can be achieved by exploiting again the property
of $d^+(x)$ and integrating by parts $s$ times, leading to:
%\bea
%{\cal E}_m^+ &=&\frac1\pi \sum_{l=m+1}^{\infty}
%\left|\frac{1}{(i l)^s}\int_0^{2\pi}dx\, e^{i x l}\frac{d}{dx^s}d^+(x)\right|
%\nonumber \\
%&\le& \frac1\pi \sum_{l=m+1}^{\infty}
%\frac{1}{l^s} \int_0^{2\pi}dx\,
%\left|\frac{d}{dx^s}d^+(x)\right|
%\nonumber \\
%&=& \frac{C_s^+}{\pi}\zeta(s,m+1),
%\eea
%
\bea
{\cal E}_m^+ &\le& \frac{C_s^+}{\pi}\zeta(s,m+1),
\eea
where $\zeta(s,m+1)$ is the generalized Riemann zeta function.
Summarizing, one has that for any integer $s$ and $m$
\begin{equation}
||\omega^+|| \le S^+_m+\frac{C_s^+}{\pi}\zeta(s,m+1)\,,
\label{wbound}
\end{equation}
the bound becoming tighter for large $m$ and $s$.

Let us now determine an upper bound for $||X||$. First, for any
finite $n$, the matrix $X$ has a bipartite symmetric structure of
the following form:
\begin{displaymath}
X=\left(
\begin{array}{cc}
0 & B \\
B & 0
\end{array}
\right)
\end{displaymath}
where $B$ is a $\frac{n}{2}\times \frac{n}{2}$ symmetric Toeplitz
matrix. Then $||X||$ coincides with $||B||$ and in order to evaluate
the latter we need to determine which row of $B$ gives rise to the
maximum in Eq.~(\ref{ninf}). Fortunately, in the macroscopic limit
$n\rightarrow\infty$ the periodic boundary conditions can be
disregarded and the last row of $B$ determines its norm. This is
because going from the last to the next to the last row we simply
remove the term $|v_1^-|$ from the sum in Eq.~(\ref{ninf}), and so on
for the other rows. It can be seen that this is the case also taking
rigorously into account the periodic boundary conditions. By writing
explicitly matrix $B$:
\begin{displaymath}
B=\left(
\begin{array}{ccccc}
v_{n/2}^- & v_{n/2-1}^- & \dots & v_2^- & v_1^- \\
v_{n/2-1}^- & \ddots & \ddots &  & v_2^- \\
\vdots & \ddots & \ddots & \ddots & \vdots \\
 v_2^- &  & \ddots & \ddots & v_{n/2-1}^- \\
 v_1^- & v_2^- & \dots & v_{n/2-1}^- & v_{n/2}^- \\
\end{array}
\right)\;.
\end{displaymath}
and exploiting its symmetries, we see that our considerations can be
restricted to the last $n/4$ rows (or, equivalently, to the first
$n/4$ rows). Define for a generic row the following sum
($j=1,\dots,n/4$ denotes the rows in reverse order, from the last to
the $\frac{n}{4}$-th one):
\begin{equation}
R_j=\sum_{l=j}^{n/2-1}|v_l^-|+\sum_{l=\frac{n}{2}+1-j}^{n/2}|v_l^-|\;,
\end{equation}
such that $||B||={\rm max}_jR_j$. The difference between $R_j$ for a
generic row and $R_1$ can be bound from above as follows:
\begin{equation}
R_j-R_1=\sum_{l=\frac{n}{2}+1-j}^{n/2}|v_l^-|-\sum_{l=1}^{j}|v_l^-|
\le\sum_{l=\frac{n}{2}+1-j}^{n/2}|v_l^-|\,,
\end{equation}
which implies that
\begin{equation}
{\rm max}_j(R_j-R_1)\le\sum_{l=\frac{n}{4}+1}^{n/2}|v_l^-|\,.
\end{equation}
Since the sum above goes to zero for $n\rightarrow\infty$ [as can be
explicitly seen by exploiting the exponential decay of $|v_l^-|$ in
\refeq{decay}], we have that the error made by considering the last
row of $B$ as the one that gives the maximum in evaluating $||B||$ can
be made as small as one may wish. In other words, we can identify
$||X||=R_1$.

Now, we can proceed as we did for $||\omega^+||$. In particular
defining again the partial sum as
\begin{equation}
S^-_m\equiv\sum_{l=1}^{m}|v_l^-|\;,
\end{equation}
we obtain that for any integer $s$ and $m$
\begin{equation}
||X|| \le S^-_m+\frac{C_s^-}{2\pi}\zeta(s,m+1)\,,
\label{xbound}
\end{equation}
where again $S^-_m$ can be calculated explicitly and the bound becomes
tighter for large $m$ and $s$.

Summarizing, considering Eqs.~(\ref{cineq}), (\ref{lmin}),
(\ref{wbound}), and (\ref{xbound}), we have shown that, in the
macroscopic limit, the log-negativity in the half-half partition
is zero when the following inequality is satisfied:
\bea
&&\!\!\!\!\!\!\!\!\!\!\!\!\!\!\!\!\!\!
2\left(S^+_m+\frac{C_s^+}{\pi}\zeta(s,m+1)\right)
\left(S^-_m+\frac{C_s^-}{2\pi}\zeta(s,m+1)\right)
\nonumber \\
&&\qquad\qquad\qquad\qquad+\left(\frac{e^{\sqrt{1+2c}/T}-1}{e^{\sqrt{1+2c}/T}+1}\right)^2
<1
\label{Thh}
\eea
Based on the formula above and on Ineq.~(\ref{negeotlim}), we depicted
in Fig.~\ref{tlim_phd} the region in the $c-T$ plane for which bound
entanglement is present in the macroscopic limit (shaded
region). We see that for any coupling $c$ we can guarantee that there
is an interval of temperatures for which the negativity in the
half-half partition is zero, nevertheless the state is entangled.

Finally, let us mention an issue concerning the procedure used in
order to derive Ineq.~(\ref{negeotlim}). We considered a finite system
composed by $n$ particles and calculated explicitly $T^{e:o}_{\rm
  th}$. We have seen that it does not depend on $n$ and so argued that
it has to be valid also in the limit $n\rightarrow\infty$. On the
other hand, one could have considered first the limit
$n\rightarrow\infty$ and then analyzed a distinguished region of the
system for increasing size [as we did in deriving Ineq.~(\ref{Thh})].
The two procedures are equivalent, since no convergence problem
appears in the spectrum of $V^{\rm n}$ for $n\rightarrow\infty$.

%We have performed numerical
%calculations by proceeding in this latter way without encountering
%any difference in $T^{e:o}_{\rm th}$. 
%%%%%%%%%%%%%%%%%%%%%%%%%%%%%%%%%%%%%%%%%%%%%%%%%%%%%%%%%%%%%%%%%%%%%%%%%
%%%%%%%%%%%%%%%%%%%%%%%%%%%%%%%%%%%%%%%%%%%%%%%%%%%%%%%%%%%%%%%%%%%%%%%%%
%%%%%%%%%%%%%%%%%%%%%%%%%%%%%%%%%%%%%%%%%%%%%%%%%%%%%%%%%%%%%%%%%%%%%%%%%

\section{Spin systems}\label{spin}
As already mentioned, we go beyond the harmonic oscillator case and
also investigate the existence of bound entangled thermal states in
spin systems. Consider the Hamiltonian
\begin{equation}\label{XY}
H_{XX}=-J\sum_{i=1}^n (\sigma^x_{i}\sigma^x_{i+1}+\sigma^y_{i}
\sigma^y_{i+1}) +B\sum_{i}^n \sigma_i^{z},
\end{equation}
that is, $n$ spin-one-half particles with nearest-neighbor $XX$
interactions with coupling constant $J$
and subject to a transverse magnetic field $B$.
% in the $z$ direction.
For this system we also calculate the thermal-state entanglement,
measured now by the negativity $E_N$, for different partitions. All
the obtained results are consistent with the previous reasoning: there
is a temperature range for which the negativity in the half-half
partitions is zero, nevertheless the system is still entangled as
proven by the negativity in the even-odd partition. Although in this
case we are not able to deal with big systems, our results suggest an
entanglement area law for non-zero temperature and the independence of
the gap $T^{e:o}_{\rm th}-T^{h:h}_{\rm th}$ with the system size, as
found in the harmonic oscillator case.
%%%%%%%%%%%%%%%%%%%%%%%%%%%%%%%%%%%%%%%%%%%%%%%%%%%%%%%%%%%%%%%%%%%%%%%%%
\begin{figure}
{\includegraphics[width=0.5\textwidth,height=0.3\textwidth]{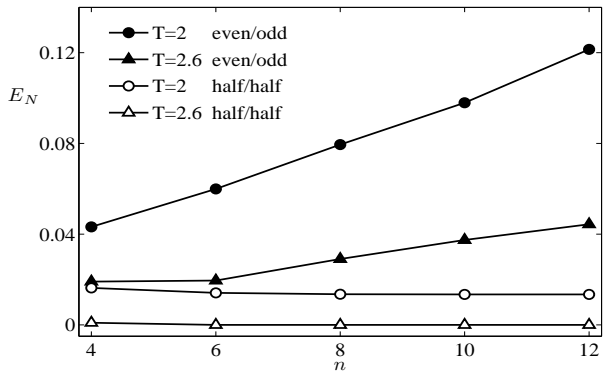}}
  \caption{Negativity in the even-odd (full symbols) and half-half
    (empty symbols) partitions for the thermal states of the
    %$n$ spin-one-half particles with
    Hamiltonian (\ref{XY}) with $B = 1.9$ and $J=1$.
    The temperatures for each partition are set to be $T =2$ and $2.6$ (see legend).
    In the even-odd partition we can clearly see
    an increase of the negativity with respect to the system size,
    whereas it saturates in the half-half partition.
    %Both behaviors are expected from the entanglement-area law.
    For $T=2.6$ we see
    that the state is bound entangled (the negativity in the half-half
    partition being zero).}
  \label{XYeo}
\end{figure}
%%%%%%%%%%%%%%%%%%%%%%%%%%%%%%%%%%%%%%%%%%%%%%%%%%%%%%%%%%%%%%%%%%%%%%%%%

In Fig.~\ref{XYeo} we plot the negativity for the even-odd (full
symbols) and for the half-half (empty symbols) partitions, for two
different values of the temperature.  For a constant number of bonds
between the partitions as in the case of the half-half splitting the
negativity does not depend on the system size $n$, whereas for the
even-odd partition it increases with $n$. This behaviour resembles the
previous study for harmonic oscillators (see Fig.~\ref{area_nn}), and
the same reasonings apply here to show the presence of bound
entanglement.

%%%%%%%%%%%%%%%%%%%%%%%%%%%%%%%%%%%%%%%%%%%%%%%%%%%%%%%%%%%%%%%%%%%%%%%%%
\begin{figure}
{\includegraphics[width=0.5\textwidth,height=0.3\textwidth]{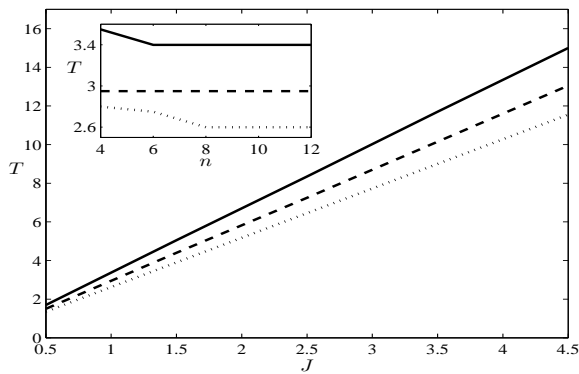}}
  \caption{Threshold temperatures above which the negativity is
    zero in the even-odd ($T^{e:o}_{\rm th}$, solid line), $1:n-1$
    ($T^{1:n-1}_{\rm th}$, dotted line) and
    half-half ($T^{h:h}_{\rm th}$, dashed line) partitions. We plot the threshold
    temperature as a function of the coupling parameter $J$
    of the Hamiltonian (\ref{XY}) with $n=10$ and $B=1.9$.
    {\bf{Inset:}} Temperature above which the negativity in the
    even-odd ($T^{e:o}_{\rm th}$, solid line), $1:n-1$
    ($T^{1:n-1}_{\rm th}$, dotted line) and half-half ($T^{h:h}_{\rm
      th}$, dashed line) partitions is zero as a function of the
    number $n$ of spins with Hamiltonian (\ref{XY}) with $J=1$ and $B=1.9$.}
  \label{fig39}
\end{figure}
%%%%%%%%%%%%%%%%%%%%%%%%%%%%%%%%%%%%%%%%%%%%%%%%%%%%%%%%%%%%%%%%%%%%%%%%%

In order to study the PPT condition in a system with $n=10$ particles for different
values of the coupling constant $J$, we plot the corresponding separability temperature
in Fig.~\ref{fig39} for three different bipartite splittings.
In all the range of values of $J$ the separability temperature
corresponding to the even-odd partition $T^{e:o}_{\rm th}$ is
strictly higher than $T^{1:n-1}_{\rm th}$ and $T^{h:h}_{\rm th}$,
corresponding to the $1:n-1$ and half-half partitions respectively.
%The even-odd partition in all the range of temperatures has a greater amount
%of entanglement than the half-half and $1:n-1$ partitions.
%The
%half-half and $1:n-1$ show similar values of the negativity
%and are equal for some $T$, which in the scope of the
%area law can be explained by the same amount of bonds that this partitions
%share between parts of the system.
The corresponding threshold temperatures for PPT are shown in the
inset of Fig.~\ref{fig39} for increasing system sizes. For $n>6$ the
gap between these temperatures appears constant (we only compute these
values for systems with an even number of particles for symmetry
reasons).

%%%%%%%%%%%%%%%%%%%%%%%%%%%%%%%%%%%%%%%%%%%%%%%%%%%%%%%%%%%%%%%%%%%%%%%%%
\begin{figure}
{\includegraphics[width=0.5\textwidth,height=0.3\textwidth]{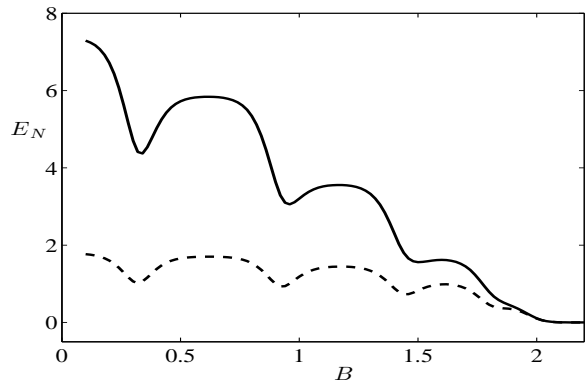}}
  \caption{Negativity in the even-odd (solid line) and half-half
    (dahed line) partitions for the thermal states of Hamiltonian (\ref{XY}) with
    $n=10$, $J=1$ and $T=0.1$
    %spin-one-half particles with  for
    as a function of
    %different values of the local field
    $B$. For $B>2$ the ground state
    is the separable state $|00\ldots0\rangle$, but
    %a small
    some amount of entanglement still appears due to the mixture with
    entangled excited states.}
  \label{fig38}
\end{figure}
%%%%%%%%%%%%%%%%%%%%%%%%%%%%%%%%%%%%%%%%%%%%%%%%%%%%%%%%%%%%%%%%%%%%%%%%%

We can tune the local field $B$ of the hamiltonian Eq.~\eqref{XY} in such a way that the the system ground state exhibits some interesting entanglement properties. For a local field $B=1.9$ and $J=1$ the ground state of the system is a generalized W-state of $n$ spins \cite{bruss}, whereas for $B>2$ the ground state is the product state $|00\ldots0\rangle$. In Fig.~\ref{fig38} we plot, for different values of the field $B$ and $T=0.1$, the negativity for the
even-odd and half-half partitions.
%For values $B<2$ the even-odd
%partition shows more entanglement than the half-half partition, with
Notice the presence of the characteristic dips due to the mixing of
different entangled ground states \cite{arnesen}. In particular, we
can see that for $B<2$ the even-odd partition has a higher degree of
entanglement than the half-half partition. As a consequence, the main
points of our intuition for the emergence of bound entanglement are
still valid, also in the presence of complex behaviors like the ones
shown in Fig.~\ref{fig38}.

\begin{figure}
{\includegraphics[width=0.5\textwidth,height=0.3\textwidth]{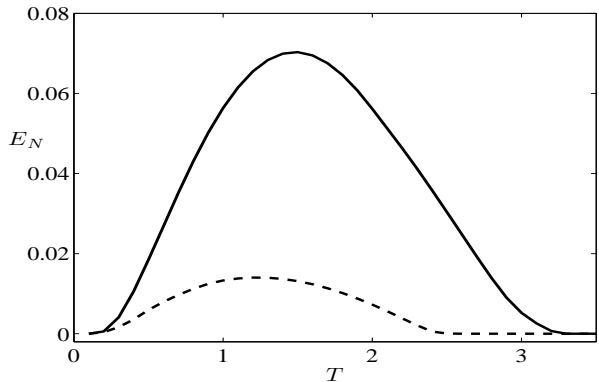}}
  \caption{Negativity in the even-odd (solid line) and half-half
    (dahed line) partitions for $n=10$, $J=1$ and $B = 2.3$ for
    spin-one-half particles with Hamiltonian (\ref{XY}).
    As expected the negativity in the even-odd partition is
    greater than the corresponding to the half-half partition,
    and bound entanglement appears in the temperature range
    $2.5<T<3.2$.}
  \label{fig35}
\end{figure}

For low temperatures and values of the local field $B>2$ the
ground state (separable) dominates the system and therefore the
thermal state is separable for all the partitions. For higher
temperatures the state turns out to be a combination of the separable
ground state and some excited states that can be entangled
\cite{berry}. In this condition entanglement can be increased by
raising the temperature.
%In this conditions some entanglement can appear for $B>2$, and by rising the temperature
%and thus incrising the degree of mixture we add more entanglement between the partitions.
This is shown in Fig.~\ref{fig35}, where we plot the negativity for
fixed field $B=2.3$ as a function of the temperature for the
even-odd and half-half partitions.
%The separable ground state
%$|00\ldots0\rangle$ produces a separable thermal state for low
%$T$.
We can see again that the negativity in the even-odd
partition is higher than the respective values for the half-half
partition. In particular, in the range $2.5<T<3.2$ the half-half
partition is PPT
%separable
while
%there is a non-negative value of
the negativity in the even-odd partition is non-null. Thus, also in
this quite peculiar conditions -- in which the ground state has
$E_N=0$ -- we see, for increasing temperatures, that {\it i)} when the
system becomes entangled (\ie, for small $T$ greater than zero) the
negativity is higher in the even-odd partition and {\it ii)} before
the system becomes separable (\ie, for higher $T$) it passes through
a non-distillable region. This behavior again supports the general
validity of our considerations.

%%%%%%%%%%%%%%%%%%%%%%%%%%%%%%%%%%%%%%%%%%%%%%%%%%%%%%%%%%%%%%%%%%%%%%%%%
%%%%%%%%%%%%%%%%%%%%%%%%%%%%%%%%%%%%%%%%%%%%%%%%%%%%%%%%%%%%%%%%%%%%%%%%%

Finally, we mention that we have found an analogous behavior in
systems subjected to the Heisenberg interaction
\begin{equation}
H_{XXX}=\sum_{i=1}^n (\sigma^x_{i}\sigma^x_{i+1}+\sigma^y_{i}
\sigma^y_{i+1} +\sigma^z_{i}\sigma^z_{i+1})+B\sum_{i}^n
\sigma_i^{z}
\end{equation}
for which a study of the entanglement properties of two
neighboring spins was shown in \cite{arnesen}. In this model a
value of the local field $B>4$ produces the ground state
$|00\ldots0\rangle$. From this point the results and
considerations for this model are essentially equivalent to those
corresponding to the XX model.

%%%%%%%%%%%%%%%%%%%%%%%%%%%%%%%%%%%%%%%%%%%%%%%%
\section{Concluding remarks}\label{esco}

In closing this paper, we point out a few remarks regarding the obtained
results. First, we have shown the existence of thermal bound
entanglement by considering local ($n$-party) distillation
strategies. One can also pose the question whether fully PPT
entanglement can exist in these systems, that is entangled states such
that all bipartite partitions are PPT. Actually, as shown in
Ref.~\cite{Geza}, fully PPT entanglement can be found for spin
systems, at least up to $9$ qubits. On the other hand, our results,
when combined with those of Refs.~\cite{AW,Hyllus}, prove that fully
PPT entangled states cannot be obtained for the harmonic systems
studied here. In Ref.~\cite{AW} bounds are derived for the threshold
temperatures for full separability of various harmonic systems.
Remarkably, these upper bounds are tight in case of systems with
sufficient translational symmetries, as in the case of the
nearest-neighbor interaction in \refeq{Vnn}. It is then possible to
depict a kind of phase diagram in the $T-c$ plane separating two
regions, one in which the state is separable and the other in which
quantum correlations are present in some form. Such a threshold turns
out to coincide with the one given by Ineq.~(\ref{negeotlim}) for the
zero negativity in the even/odd partition. Thus, the thermal states of
this system become PPT and fully separable for the same temperature.
We mention here that such a result holds also for other systems, for
example for the one given in \refeq{Vuf}. However, in these cases we had
to approach the problem numerically (based on the results of
Ref.~\cite{Hyllus}) and consider only finite systems.

Finally, let us stress that our results enrich the above mentioned
phase diagram of Ref.~\cite{AW}, in the sense that we can distinguish
a third region in it where quantum correlations are present but they
are not distillable. In particular, referring to Figs.~\ref{PhD} and
\ref{tlim_phd}, we have that, by rising the temperature at fixed
coupling, the system goes from a genuinely entangled state (in which
all the bipartitions are NPPT) to a bound entangled state and finally
becomes just classically correlated. As already discussed, we have
found that such a behavior is common to all the harmonic systems we
took in consideration. For spin systems, a similar phase diagram
holds, at least for the finite-size systems studied here.

In conclusion, we have shown the existence of thermal-state bound
entanglement for several quantum many-body systems. We explicitly
considered systems composed by hundreds of oscillators and observed in
general an area law for thermal states. This findings turned out to be
crucial in proving the existence of bound entanglement also in the
macroscopic limit for harmonic systems. As a consequence, and by virtue of
the generality of area laws, these results have been shown to
be valid for a variety of other systems, as well as in the presence of
complex behaviors of the entanglement. All these considerations strongly
support the existence of thermal-state bound entanglement in the
macroscopic limit also for finite-dimensional systems.

\begin{acknowledgements}
  We thank S. M. Giampaolo, F. Illuminati, G. Toth and M. O. Terra Cunha for helpful
  comments. This work is supported by the EU QAP project, the Spanish MEC, under
  FIS2007-60182 and Consolider-Ingenio QOIT projects, and a ``Juan de
  la Cierva" grant, the Generalitat de Catalunya, and the Universit\'a
  di Milano under grant ``Borse di perfezionamento all'estero".
\end{acknowledgements}


\begin{thebibliography}{30}
\bibitem{note1} In fact, due to the translational invariance of the
  systems considered here, all the half-half partitions are
  equivalent.
\bibitem{note2} Notice that the details of the model (in particular,
  the peculiar independence of the coupling between next to nearest
  neighbors on $\mu$) implies that for small $\mu$ the negativity is
  higher in the contiguous partitions then in the even-odd one.
\end{thebibliography}
\end{document}